\documentclass[preprint]{elsarticle}

\usepackage{amsmath}
\usepackage{booktabs}
\usepackage{hepnames}
\usepackage{hyperref}
\usepackage{mathtools}
\usepackage{multirow}
\usepackage{physics}
\usepackage[separate-uncertainty=true]{siunitx}
\usepackage{standalone}
\usepackage{subcaption}
\usepackage{textcomp}
\usepackage{tikz}
\usepackage{xpatch}

\usepackage{lineno}
\makeatletter
\xpatchcmd\@HepConStyle
 {\edef\@upcode{\updefault}}
 {\ifdefined\shapedefault\edef\@upcode{\shapedefault}\else\edef\@upcode{\updefault}\fi}
 {}{}
\makeatother
\journal{Nuclear Instruments and Methods in Physics Research A}

\biboptions{numbers,sort&compress}
\newcommand{\sub}[1]{\ensuremath{_{\text{#1}}}}
\newcommand{\supp}[1]{\ensuremath{^{\text{#1}}}}

\begin{document}

%\linenumbers

%%%%%%%%%%%%%%%%%%%%%%%%%
%%%%%%%%%%%%%%%%%%%%%%%%%
\begin{frontmatter}

  \title{Measurement of ionization quenching in plastic scintillators}

  %% authors
  \author{Thomas P\"oschl\corref{mycorrespondingauthor}}
  \cortext[mycorrespondingauthor]{Corresponding author}
  \ead{thomas.poeschl@ph.tum.de}
  \author{Daniel Greenwald}
  \author{Martin J. Losekamm}
  \author{Stephan Paul}
  \address{%
    Technical University of Munich,
    Department of Physics,
    Institute for Hadronic Structure and Fundamental Symmetries,
    James-Franck-Str. 1,
    D-85748 Garching, Germany%
  }

  %%%%%%%%%%%%%%%%%%%%%%%%%
  \begin{abstract}

    Plastic scintillators are widely used in high-energy and medical
    physics, often for measuring the energy of ionizing
    radiation. Their main disadvantage is their non-linear response to
    highly ionizing radiation, called ionization quenching. This
    nonlinearity must be modeled and corrected for in applications
    where an accurate energy measurement is required. We present a new
    experimental technique to granularly measure the dependence of
    quenching on energy-deposition density. Based on this method, we
    determine the parameters for four commonly used quenching models
    for two commonly used plastic scintillators using protons with
    energies of \SI{30}{MeV} to \SI{100}{MeV}; and compare the models
    using a Bayesian approach. We also report the first
    model-independent measurement of the dependence of ionization
    quenching on energy-deposition density, providing a purely
    empirical view into quenching.
    
  \end{abstract}
  %%%%%%%%%%%%%%%%%%%%%%%%%

  %%%%%%%%%%%%%%%%%%%%%%%%%
  %%%%%%%%%%%%%%%%%%%%%%%%%

\end{frontmatter}
%%%%%%%%%%%%%%%%%%%%%%%%%
%%%%%%%%%%%%%%%%%%%%%%%%%

%%%%%%%%%%%%%%%%%%%%%%%%%
%%%%%%%%%%%%%%%%%%%%%%%%%
\section{Introduction}

Plastic scintillators are widely used in particle detectors in
high-energy physics experiments and for medical applications, for
example in radiotherapy~\cite{Kharzheev2015,Beaulieu2016}. When a
charged particle passes through a scintillator, it loses energy by
ionizing scintillator molecules, leaving them in excited states. These
excited molecules can emit light when they relax, providing a
measurable signal. Since the light yield is related to a particle's
energy loss, scintillators are used in calorimeters.  Since
scintillators are easily segmented, they are also used in particle
trackers.

When the density of energy lost by a particle is low, a scintillator's
light yield is linearly proportional to the energy lost. When the
density is high, the production of detectable light is hindered by
various processes and the light yield is nonlinearly proportional to
the energy lost~\cite{Birks1951}. This reduction is called ionization
quenching~\cite{Brooks1979}.

At high momentum, a particle loses relatively little energy per
distance in comparison to its kinetic energy; so its momentum
decreases very slowly. The energy-loss rate is also only weakly
dependent on the particle's momentum; so the light yield per distance
at high momentum is nearly constant and easily calibrated. This means
scintillators produce consistent and reliable signals useful for
tracking high-energy particles.

Additionally, if a particle is stopped (or significantly slowed)
within a scintillator, the light produced can be used to measure the
particle's energy. But at low momentum, a particle loses large amounts
of energy over short distances, so ionization quenching significantly
reduces the light emission and complicates the energy
measurement. Accurate calorimetry of highly ionizing radiation
requires a good understanding of quenching.

The magnitude of quenching is dependent on the amount of energy
deposited in a scintillator per unit distance. Many competing models
parametrize this dependence via quenching
functions~\cite{Birks1951,Chou1952,Craun1970,Wright1953,Voltz1966}. Most
of these functions are based on assumptions about discrete
interactions of particles with scintillator molecules and of excited
scintillator molecules with each other. But none are capable of
predicting the values of their parameters---they are empirical.

Using data collected with two widely-used plastic scintillators, we
determine the parameters for four quenching
models~\cite{Birks1951,Chou1952,Wright1953,Voltz1966}. Using Bayesian
statistics, we compare the probabilities for the models to explain our
data. We also fit a model-independent quenching function to our data
to learn about the dependence of quenching on energy deposition free
from assumptions. We compare this result to those of the four models
we tested and discuss their advantages and deficiencies. To our
knowledge, this is the first direct measurement of a quenching
function without modeling.

%%%%%%%%%%%%%%%%%%%%%%%%%
%%%%%%%%%%%%%%%%%%%%%%%%%
\section{Quenching models}
 
The response of a scintillating material to a charged particle is
characterized by its light yield per unit of distance traveled by the
particle, $\dv*{L}{x}$. However, the light yield per unit of energy
deposited by the particle over that distance, $\dv*{L}{E}$, is more
useful in modeling and simulation. The two are related to each other
by
\begin{equation}
  \dv{L}{x} = \dv{L}{E} \cdot \dv{E}{x},
  \label{eqn:dLdx_dLdE}
\end{equation}
where $\dv*{E}{x}$ is the energy lost by the particle per unit of
distance, which depends on the energy of the particle (as well as its
species and the scintillator material). To simplify our equations, we
denote $\dv*{E}{x}$, a function of the particle's kinetic energy, $T$,
as $\epsilon(T)$. The light yield per unit of energy deposited is a
function of $\epsilon$, and therefore indirectly of $T$:
\begin{equation}
  \eval{\dv{L}{E}}_T \equiv S \cdot Q\qty(\epsilon(T)),
  \label{eqn:dLdE_SQ}
\end{equation}
where $Q(\epsilon)$ is the unitless quenching function, defined such
that at small $\epsilon$ (that is, at high kinetic energy), it goes to
unity; $S$ is the linear proportionality of light yield to energy
deposition at high energy and has units of photons per energy. So the
light yield per unit of distance is
\begin{equation}
  \eval{\dv{L}{x}}_T = \epsilon(T) \cdot S \cdot Q\qty(\epsilon(T)).
  \label{eqn:dLdx_SQeps}
\end{equation}

Scintillation light is produced via several steps~\cite{Birks1951}: a
passing particle ionizes molecules of the scintillator's base plastic
material, which then emit light. In a pure plastic, this light is
quickly reabsorbed by other molecules. To allow the light to propagate
further, the plastic is doped with a molecule that absorbs this light
and emits light of a shifted wavelength. Since neither the base
molecule nor the dopant efficiently absorbs the wave-length-shifted
light, it propagates long distances. However, dopant molecules can
absorb photons without re-emitting them or can re-emit them at
wavelengths unsuitable for detection. This occurs when they have been
excited by interaction with the ionizing particle.

J.~B.~Birks developed the first model of ionization quenching in the
early 1950s, which is still widely used~\cite{Birks1951}. He
parametrized quenching in terms of the density of excited dopant
molecules, $B$, and the probability for non-radiative relaxation, $k$:
\begin{equation}
  Q\sub{Birks}(\epsilon) = 1 / (1 + kB \, \epsilon).
  \label{eqn:Birks}
\end{equation}
Since $k$ and $B$ appear only as a product, they act as one parameter,
$kB$, called Birks' coefficient, which has units of distance per
energy. Its value depends on the scintillating material.

Many authors have extended Birks' model: Chou~\emph{et al.}\ accounted
for secondary effects by adding a term to the denominator that is
second order in $\epsilon$:
\begin{equation}
  Q\sub{Chou}(\epsilon) = 1 / (1 + kB \, \epsilon + C \epsilon^2),
  \label{eqn:Chou}
\end{equation}
where $\sqrt{C}$ has the same units as $kB$~\cite{Chou1952,Craun1970}.
Wright~\emph{et al.}\ defined the phenomenological quenching function
\begin{equation}
  Q\sub{Wright}(\epsilon) \equiv \frac{1}{W \epsilon} \log(1 + W \epsilon),
  \label{eqn:Wright}
\end{equation}
where $W$ has the same units as $kB$~\cite{Wright1953}. Voltz~\emph{et
  al.}~developed the first model to distinguish between primary and
secondary ionization: The primary particle can produce high-energy
electrons as it ionizes the scintillator molecules. They travel away
from the path of the primary particle, losing energy via ionization of
the scintillator and spreading out the energy deposition, which
weakens quenching. The Voltz model assumes that a fraction of
deposited energy, $f$, is unquenched and parametrizes the quenching of
the remaining fraction with an exponential function:
\begin{equation}
  Q\sub{Voltz}(\epsilon) \equiv f  + (1-f) e^{-V(1-f)\epsilon},
  \label{eqn:Voltz}
\end{equation}
where $V$ has the same units as $kB$~\cite{Voltz1966}.

Like Birks' coefficient, $C$, $W$, and $V$ all depend on the
scintillator material. All four must be positive and are
independent of the species of the particle interacting with the
scintillator. The Voltz model's $f$ depends on both the scintillator
material and primary particle species~\cite{Rossi1952}. None of the
parameters can be predicted from first principles---all must be
measured experimentally.

These quenching functions have some common features: As we require of
a quenching function, they are all bounded by 1 above, which is
approached as $\epsilon\to0$; and by 0 below, which may be approached
as $\epsilon\to\infty$. All have negative first
derivatives~($\dv*{Q}{\epsilon}$) everywhere regardless of their
parameters and therefore always monotonically decrease. Birks',
Wright's, and Voltz' functions all have positive second derivatives
everywhere regardless of their parameter values; only Chou's function
allows for a negative second derivative and a potential inflection
point. These properties will be important when we compare
model-dependent and model-independent results.

%%%%%%%%%%%%%%%%%%%%%%%%%
%%%%%%%%%%%%%%%%%%%%%%%%%
\section{Quenching measurement}

To determine each model's parameters and which model most accurately
describes quenching, we measure $\dv*{L}{E}$ at several kinetic
energies and fit the parameterizations of $Q(\epsilon)$ to this data
using equation~(\ref{eqn:dLdE_SQ}).

Many issues complicate this task: We cannot directly measure
$\dv*{L}{E}$; instead we measure the amount of light, $L$, produced by
a particle that has lost energy in the scintillator. So we must
integrate equation~(\ref{eqn:dLdE_SQ}):
\begin{equation}
  L(T\supp{in}, T\supp{out}) = S \! \int\limits_{\hspace{-1em}\mathrlap{T\supp{out}}}^{\mathrlap{T\supp{in}}} \! Q\qty(\epsilon(T)) \, \dd{T},
  \label{eqn:int_dLdE_SQ_dT}
\end{equation}
where $T\supp{in}$ and $T\supp{out}$ are the incoming and outgoing
kinetic energies of the particle. $Q$ is not directly a function of
$T$, but instead of $\epsilon(T)$, which is a stochastic function: At
a particular kinetic energy, we know the mean energy loss per unit
distance for particles with that energy from both the Bethe formula
and experiment~\cite{PDG2018, ICRU49}. But an individual particle's
energy loss stochastically deviates from the mean according to
distributions whose shapes are also $T$ dependent~\cite{ICRU49,
  Landau:1944, Vavilov:1957}. This stochastic behavior is difficult
and computationally expensive to model. So instead of studying the
behavior of individual particles, we study the behavior of an ensemble
of particles. We measure the distribution of $L(T\supp{in},
T\supp{out})$ and fit the quenching model parameters to the mean
amount of light, $\bar{L}$, produced by an ensemble of particles given
a mean energy loss, $\bar\epsilon$:
\begin{equation}
  \bar{L}(\bar{T}\supp{in}, \bar{T}\supp{out}) = S \! \int\limits_{\hspace{-1em}\mathrlap{\bar{T}\supp{out}}}^{\mathrlap{\bar{T}\supp{in}}} \! \bar{Q}\qty(\bar{\epsilon}(T)) \, \dd{T},
  \label{eqn:int_dLdE_SQ_dT_barred}
\end{equation}
where $\bar{T}\supp{in}$ and $\bar{T}\supp{out}$ are the mean incoming
and outgoing kinetic energies of the ensemble and $\bar{Q}$ is the
quenching function of the mean energy of an ensemble. We assume
quenching of the mean energy loss is described identically to
quenching of the stochastic energy loss: $\bar{Q} = Q$.

The above equations are further complicated by how the scintillation
light is measured: it propagates through the scintillator to a light
detector. Both propagation and detection cause losses of light. In our
experimental setup, these losses linearly scale the light yield and
can be canceled out by measuring with respect to a reference light
yield. To simplify our calculations, we measure with respect to the
signal produced by a particle with $\epsilon x \ll T$ for distances,
$x$, even much larger than our setup. To very good approximation, the
light yield of such a particle is
\begin{equation}
  \bar{L}\sub{ref} = S \! \int\limits_{\mathrlap{\hspace{-0.6em}\bar{T} - \epsilon\sub{ref} \bar{x}}}^{\mathrlap{\bar{T}}} \! \bar{Q}\qty(\bar\epsilon(T)) \, \dd{T}
  \approx \bar\epsilon\sub{ref} \, \bar{x} \, S \, \bar{Q}(\bar\epsilon\sub{ref}),
\end{equation}
where $\bar\epsilon\sub{ref}$ is the mean energy loss per unit
distance of the reference particle and $\bar{x}$ is the mean length of
scintillator passed through. We define the relative mean light yield
as
\begin{equation}
  \bar\Lambda(\bar{T}\supp{in}, \bar{T}\supp{out}) \equiv \frac{\bar{L}(\bar{T}\supp{in}, \bar{T}\supp{out})}{\bar{L}\sub{ref}}
  = \frac{1}{\bar\epsilon\sub{ref} \, \bar{x} \, \bar{Q}\qty(\bar\epsilon\sub{ref})} \, \int\limits_{\hspace{-1em}\mathrlap{\bar{T}\supp{out}}}^{\mathrlap{\bar{T}\supp{in}}} \! \bar{Q}\qty(\bar{\epsilon}(T)) \, \dd{T.}
  \label{eqn:Lambda_Q_relation}
\end{equation}

To gather granular data for a range of $\bar\epsilon$, we use a
segmented detector consisting of an array of scintillating fibers laid
in a row. We shoot a beam of protons and pions into the detector such
that pions could traverse all fibers successively and protons could
stop within the array. We vary the energies of the beams and the angle
of incidence on the fibers, $\theta$, which changes $\bar{x}$. The
protons serve as test particles for measuring quenching, and the pions
serve as the low-$\epsilon$ reference particles for the relative light
yield measurement. From initial kinetic energies in the range of tens
to hundreds of \si{MeV} to stopping, the range of $\bar\epsilon$ for
the protons varies by two orders of magnitude, while the through-going
pions are always in their minimum-ionizing energy range, regardless of
incoming beam energy. Though changes of the angle only translate into
small changes of the protons' path lengths in the fibers, they
strongly affect their energy-loss profiles because of the high
stopping power of protons shortly before stopping. Where the energy
loss is largest, quenching effects are most pronounced; so small
variations of the angle lead to large variations in quenching and
significantly improve our measurement sensitivity. We label each
different setting of beam energy and incidence angle as a run.

In each run, we measure $\bar\Lambda_i$ for each fiber, with $i$
labeling the fiber; this is the data set for each run. Unfortunately
we do not know $\bar{T}\supp{in}_i$ and $\bar{T}\supp{out}_i$ for
individual fibers. In our fits to the $\bar\Lambda_i$, the mean energy
of the proton beam prior to it entering the fiber array, $\bar{T}_0$
is a free parameter. We calculate all the incoming and outgoing
energies, $\bar{T}\supp{in}_i$ and $\bar{T}\supp{out}_i$ ($i \ge 1$),
with the continuous-slowing-down approximation~(CSDA) using data from
the National Institute of Standards and Technology~(NIST) for
$\bar{\epsilon}(T)$~\cite{ICRU1956, Pstar2005}. We account for
inactive coatings on the fibers, so $\bar{T}_0 \ne \bar{T}\supp{in}_1$
and $\bar{T}\supp{out}_i \ne \bar{T}\supp{in}_{i+1}$. This calculation
depends on $\bar{x}$, and therefore on the incidence angle, which is
also a free parameter in our fits.

%%%%%%%%%%%%%%%%%%%%%%%%%
\subsection{Fit likelihood}

To fit a quenching model, $M$, to our data, we must quantify how well
it describes the data given particular values of its parameters,
$\vec\lambda$. This is the likelihood of the data given the model and
its parameters. Since the likelihood for the data of an individual run
has a common form for all runs, we factorize the likelihood to
describe our total data set, $\vec{D}$, into the product of
likelihoods to describe the data of individual runs, $\vec{D}_r$:
\begin{equation}
  \mathcal{L}(\vec{D} \,|\, \vec\lambda, \vec\nu; M)
  \equiv
  \prod_r \mathcal{L}_r (\vec{D}_r \,|\, \vec\lambda, \vec\nu_r; M),
  \label{eqn:total_likelihood}
\end{equation}
where $\vec\nu$ is the vector of beam parameters vectors, $\vec\nu_r =
\{\bar{T}_{r0}, \theta_r\}$, for all runs; and the data for a
particular run, $\vec{D}_r$, are the observed
$\bar\Lambda_{ri}\supp{obs}$ and their uncertainties,
$\sigma_{ri}$. The likelihood for an individual run is
\begin{equation}
  \mathcal{L}_r(\vec{D}_r \,|\, \vec\lambda, \vec\nu_r; M)
  \equiv
  \prod_i \mathcal{N}\qty(\bar\Lambda_{i}\supp{exp}(\vec\lambda, \vec\nu_r; M) \,|\, \bar\Lambda_{ri}\supp{obs}, \sigma_{ri}),
  \label{eqn:run_likelihood}
\end{equation}
where $\mathcal{N}$ is the normal distribution and
$\bar\Lambda_i\supp{exp}$ is the expectation for the quenched mean
relative light yield calculated according to
equation~(\ref{eqn:Lambda_Q_relation}) using quenching model $M$.

To calculate $\bar\Lambda_i\supp{exp}$, we first simulate the
trajectory of a particle with an initial energy $\bar{T}_{r0}$ and
incidence angle $\theta_r$ through the fiber array, calculating its
energy losses in both the active and inactive layers of the array with
NIST's CSDA data. From the simulation we know the integration limits
of equation~(\ref{eqn:Lambda_Q_relation}). To account for the
variation in the pion momentum from run to run, we replace
$\bar\epsilon\sub{ref}$ in equation~(\ref{eqn:Lambda_Q_relation}),
with a run-dependent mean pion energy-loss density,
$\bar\epsilon^{\!\;\Ppi}_r$. The mean distance traversed by a pion in
a fiber is calculated from the incidence angle for the run: $\bar{x} =
\flatfrac{w}{\cos\theta_r}$, where $w$ is the width of the active
layer of a fiber. To emphasize the parameter dependence, we rewrite
equation~(\ref{eqn:Lambda_Q_relation}) explicitly for this context:
\begin{equation}
  \bar\Lambda_i\supp{exp}(\vec\lambda, \vec\nu_r; M)
  = \frac{\cos\theta_r}{\bar\epsilon^{\!\;\Ppi}_r w \; \bar{Q}_M(\bar\epsilon^{\!\;\Ppi}_r\!; \vec\lambda)} \;
  \int\limits_{\hspace{-1em}\mathrlap{\bar{T}\supp{out}_i(T_{r0}, \theta_r)}}^{\mathrlap{\bar{T}\supp{in}_i(T_{r0}, \theta_r)}} \! \bar{Q}_M\qty(\bar{\epsilon}(T); \vec\lambda) \dd{T},
\end{equation}
where now $\vec\nu_r = \{\bar{T}_{r0}, \theta_r,
\bar\epsilon^{\!\;\Ppi}_r\}$.

Runs with different angles were taken at common beam momenta; and runs
with different beam momenta were taken at common angles. Runs with a
common beam momentum share a single $T_{r0}$ and a single
$\bar\epsilon^{\,\Ppi}_r$; and runs with a common angle share a single
$\theta_r$.

We explore the parameter space of each model using a Bayesian
formulation of probability and a Markov-Chain Monte-Carlo (MCMC)
algorithm implemented by the Bayesian Analysis
Toolkit~\cite{Caldwell2009,Beaujean:2015bwl,BAT2018}. This defines the
posterior probability---the probability for parameters given our
knowledge after the experiment---as the product of the above
likelihood and a prior probability:
\begin{equation}
  P(\vec\lambda, \vec\nu \,|\, \vec{D}; M) \propto \mathcal{L}(\vec{D} \,|\, \vec\lambda, \vec\nu; M) \times P_0(\vec\lambda, \vec\nu \,|\, M),
  \label{eqn:posterior_prob}
\end{equation}
where proportionality is used since the right-hand side must be
normalized for the product to be a probability. The prior
probability, $P_0$, of parameters reflects our knowledge before the
experiment. For each model and for each scintillator type, we fit the
parameters to all data sets simultaneously. The free parameters in
each fit are all $T_{r0}$, $\theta_r$, and
$\bar\epsilon^{\!\;\Ppi}_r$ and the parameters of the quenching model
studied.

This approach necessitates that we choose a prior probability
distribution for all parameters. Although we have precise knowledge of
the proton and pion energies in the beam, we use informative uniform
prior probability distributions for the $T_{r0}$ and
$\bar\epsilon^{\!\;\Ppi}_r$. We do this since the beam passes through
two windows and a short gap of air before entering the detector
array. Interaction with the windows and air smears out the energy
distribution. The prior for each $\theta_r$ is a normal distribution
with mean and standard deviation learned from an independent fit to
pion data that calibrates the experiment's rotatory table. The priors
for the model parameters are discussed below alongside the fit
results.

The NIST data used to calculate the $\bar\Lambda\supp{exp}$ has an
uncertainty that scales the entire stopping-power data set up or down
together, not affecting the $T$ dependence of $\bar\epsilon$. We
account for this uncertainty with a parameter that scales the CSDA
data. It has a normal prior probability distribution centered at unity
with a standard deviation of \SI{4}{\percent}---the known uncertainty
from NIST. This parameter is also free in the fit, but its posterior
probability is identical to its prior probability. Though this
uncertainty affects all analyses that rely on NIST data, it has been
neglected in most existing measurements.

%%%%%%%%%%%%%%%%%%%%%%%%%
\subsection{Model comparison}

We compare models to each other by calculating Bayes factors, which
quantify the relative abilities of two models to describe the data
regardless of the best-fit values found for their
parameters~\cite{Kass1995}. This approach accounts for model
complexities, full posterior probability distributions, and
overfitting (acting as an Occam's razor).

The Bayes factor, $K\sub{AB}$, comparing model A to model B, is the
ratio of the model evidences, $z\sub{A}$ and $z\sub{B}$,
\begin{equation}
  K\sub{AB} \equiv \frac{z\sub{A}}{z\sub{B}}.
\end{equation}
The evidence of a model is a measure of its ability to describe the
data regardless of the values of its parameters. It is the integral
over the right-hand side of equation~\ref{eqn:posterior_prob}
\begin{equation}
  z_M = \int \! \mathcal{L} (\vec{D} \,|\, \vec\lambda, \vec\nu; M) \,
  P_0(\vec\lambda,\vec\nu \,|\, M) \, \dd\vec\lambda \, \dd\vec\nu,
  \label{eqn:evidence}
\end{equation}
---the integrand is the product of the likelihood and the prior
probability density and the integration is over all parameters and
over the entirety of each parameter's allowed range.

The posterior belief in preferring model A over model B is
\begin{equation}
  \frac{P(\text{A})}{P(\text{B})} = K\sub{AB} \cdot \frac{P_0(\text{A})}{P_0(\text{B})},
\end{equation}
where $P_0(M)$ and $P(M)$ are the prior and posterior probabilities
for a particular model---that is, one's belief in the model before and
after the experiment. The prior probabilities, $P_0(A)$ and $P_0(B)$,
are subjectively chosen by each scientist. The Bayes factor thus
quantifies the objective part of our learning process and separates it
from the subjective priors. If $K\sub{AB}$ is greater than one, model
A is preferred over model B by the data; if $K\sub{AB}$ is less than
one, model B is preferred over model A by the data.

The integral in equation~(\ref{eqn:evidence}) is not generally easy to
calculate. We used a harmonic-mean estimator~(HME) algorithm to
calculate evidences from the MCMC samples~\cite{Newton1994}. We
calculate the evidence from the samples by
\begin{equation}
  z\sub{M} \approx \qty(\frac1N \sum_{i = 1}^{N} \frac{1}{
  \mathcal{L} (\vec{D} \,|\, \vec\lambda_i, \vec\nu_i; M) \, P_0(\vec\lambda_i,\vec\nu_i \,|\, M)
  })^{\!\!\!-1},
\end{equation}
where the sum is over the $N$ sampled parameter points in the Markov
chain. Since this method suffers from numerical instabilities in
regions of small posterior probability density, we restricted our
evaluation of the HME to a volume in which the calculation is well
behaved and accounted for this restriction in calculating the evidence
using an algorithm developed in~\cite{Caldwell2018}.

%%%%%%%%%%%%%%%%%%%%%%%%%
%%%%%%%%%%%%%%%%%%%%%%%%%
\section{Experimental setup}

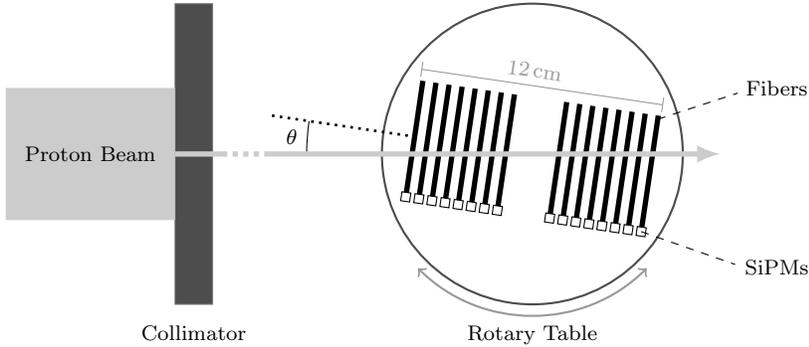
\begin{figure}[t!]
  \begin{tikzpicture}[scale=0.5, font=\footnotesize]
  \clip (-5.5,-5.5) rectangle (17,4.5);
  
  \begin{scope}[shift={(9,0)}]

    %% rotary table
    \draw[color=black!70, thick] (0,0) circle (4);
    \draw[color=black!40, thick, <->] (-3.04, -3.04) arc (-135:-45:4.3);
    \node[align=center,below] at (0,-4.3) {Rotary Table};

    %% fibers
    \def\tablerotation{-8.4:(0,0)}
    \foreach \x in {-90,-80,-70,-60,-50,-40,-30,-20,20,30,40,50,60,70,80,90} {
      \draw[color=black, line width=2, rotate around=\tablerotation, xshift=\x] (0,-1.5) -- +(0, +3);
      \draw[color=black, rotate around=\tablerotation, xshift=\x] (-0.12,-1.5) rectangle +(0.24,-0.24);
    }
    \draw[color=black!40, rotate around=\tablerotation, |-|] (-3.25,1.8) -- +(6.5,0);
    \node[color=black!40, align=center, above, rotate around=\tablerotation] at (0,1.8) {\SI{12}{cm}};
    
    %% fiber label
    \draw[color=black, rotate around=\tablerotation, xshift=90, thin, dashed] (0,1.4) -- +(2,1);
    \node[color=black] at (6.5,1.75) {Fibers};
    
    %% SiPM label
    \draw[color=black, rotate around=\tablerotation, xshift=90, thin, dashed] (0,-1.62) -- +(2.6,-0.5);
    \node[color=black] at (6.5,-3) {SiPMs};
    
    %% angle arc
    \draw[color=black, line width=1, dotted, rotate around=\tablerotation] (-7,0) -- (-3,0);
    \draw[color=black] (-6,0) arc (0:-8.4:-6);
    \node[align=right, left] at (-6,0.45) {$\theta$};
    
  \end{scope}

  %% Collimator
  \filldraw[color=black!50, fill=black!70, ultra thin] (+0.5,-4) -- (+0.5,+4) -- (-0.5,+4) -- (-0.5,-4) -- cycle;
  \node[align=center,below] at (0,-4.3) {Collimator};
  
  %% beam line
  \draw[color=black!20, line width=2] (-0.5,0) -- (0.75,0);
  \draw[color=black!20, line width=2, dotted] (0.75,0) -- (2,0);
  \draw[color=black!20, line width=2, ->, >=latex] (2,0) -- (14,0);
  \draw[color=black!20, line width=50] (-5,0) -- (-0.5,0);
  \node [align=center] at (-2.75,0) {Proton Beam};
  
\end{tikzpicture}
	\caption{Schematic of the experimental setup.\label{fig:experimentalSetup}}
\end{figure}

Our detector consists of 16 scintillating fibers, each \SI{71}{mm}
long with a square $\SI{2}{mm}\times \SI{2}{mm}$ cross section. We arrange
them such that their long sides were perpendicular to the beam and
place them in a row such that the beam passed through them
sequentially. Figure~\ref{fig:experimentalSetup} shows a schematic
view of the experimental setup. We measured with two different
scintillating materials: SCSF-78 from Kuraray, with a polystyrene
base; and BC-408 from Saint-Gobain, with a polyvinyltoluene
base~\cite{Kuraray2018, Saint-Gobain2018}.

The light produced in each fiber is detected by a square $\SI{3}{mm}\times\SI{3}{mm}$ Hamamatsu Photonics S13360-4935 silicon
photomultiplier (SiPM) glued to one end of the
fiber~\cite{Hamamatsu2018}. Each SiPM has a pitch size of
\SI{25}{\micro m} with \num{14400} pixels in total, of which \SI{6400}
overlap with the fiber end. The large SiPM eases the gluing process and minimizes variations due to positioning errors. The measurements were performed with constant overvoltages on the SiPMs and constant temperatures to ensure consistent gains throughout measuring. From simulation, we estimate an average
SiPM signal of 10 to 15 photoelectrons for a pion and around 200
photoelectrons for the maximum signal from a proton. So saturation
effects are negligible and we have constant light detection efficiency
for both the pions and the protons~\cite{Renker2006}. From test measurements in which we varied the vertical position of the detector relative to the beam, we observed no dependence of our measurements on this alignment. To digitize the SiPM signals, we use multichannel mezzanine-sampling analog-to-digital converters~(ADCs)~\cite{Mann2009}.

We measured in the \Ppi{}M1 beamline of the high-intensity proton
accelerator at the Paul Scherrer Institute~\cite{Psi2018}. The
\Ppi{}M1 beam consists of protons and pions with momenta adjustable
between \SI{220}{MeV/c} and \SI{450}{MeV/c} with a resolution of about
\SI{1}{\percent}~\cite{Psi2017}. The beam spot size was
$\SI{10}{mm}\times\SI{10}{mm}$~(at fwhm) and centered on the middle of
our fibers. To reduce the beam divergence, we placed a copper
collimator with a \SI{2}{mm}-diameter bore before the fiber array,
with \SI{20}{cm} between the exit of the collimator and the first
fiber (at perpendicular incidence). The collimator produced a strongly
collimated beam of protons that hit the center of the fiber array (at
perpendicular incidence), but did not significantly alter the pion
beam. The fiber array was mounted on a rotary table that allowed us to
vary the angle of incidence of the beam on the array. The entire setup
was placed in a vacuum chamber to minimize beam interactions with air
before entering the detector.

The recorded data set for the SCSF-78 scintillator contains seven
runs: five with an incidence angle of \SI{1.6}{\degree} at momenta of
\SI{230}{MeV/c}, \SI{240}{MeV/c}, \SI{275}{MeV/c}, \SI{300}{MeV/c},
and \SI{335}{MeV/c}; and two further at \SI{3.4}{\degree} and
\SI{8.4}{\degree}, both at \SI{335}{MeV/c}. The recorded data set for
the BC-408 scintillator contains six runs: four with an incidence
angle of \SI{1.6}{\degree} at momenta of \SI{240}{MeV/c},
\SI{300}{MeV/c}, \SI{335}{MeV/c}, and \SI{350}{MeV/c}; and two further
at \SI{5.4}{\degree} and \SI{7.9}{\degree}, both at \SI{335}{MeV/c}.

%%%%%%%%%%%%%%%%%%%%%%%%%
\subsection{Relative light yield measurement}

\begin{figure}[t!]
	\includegraphics[width=\linewidth]{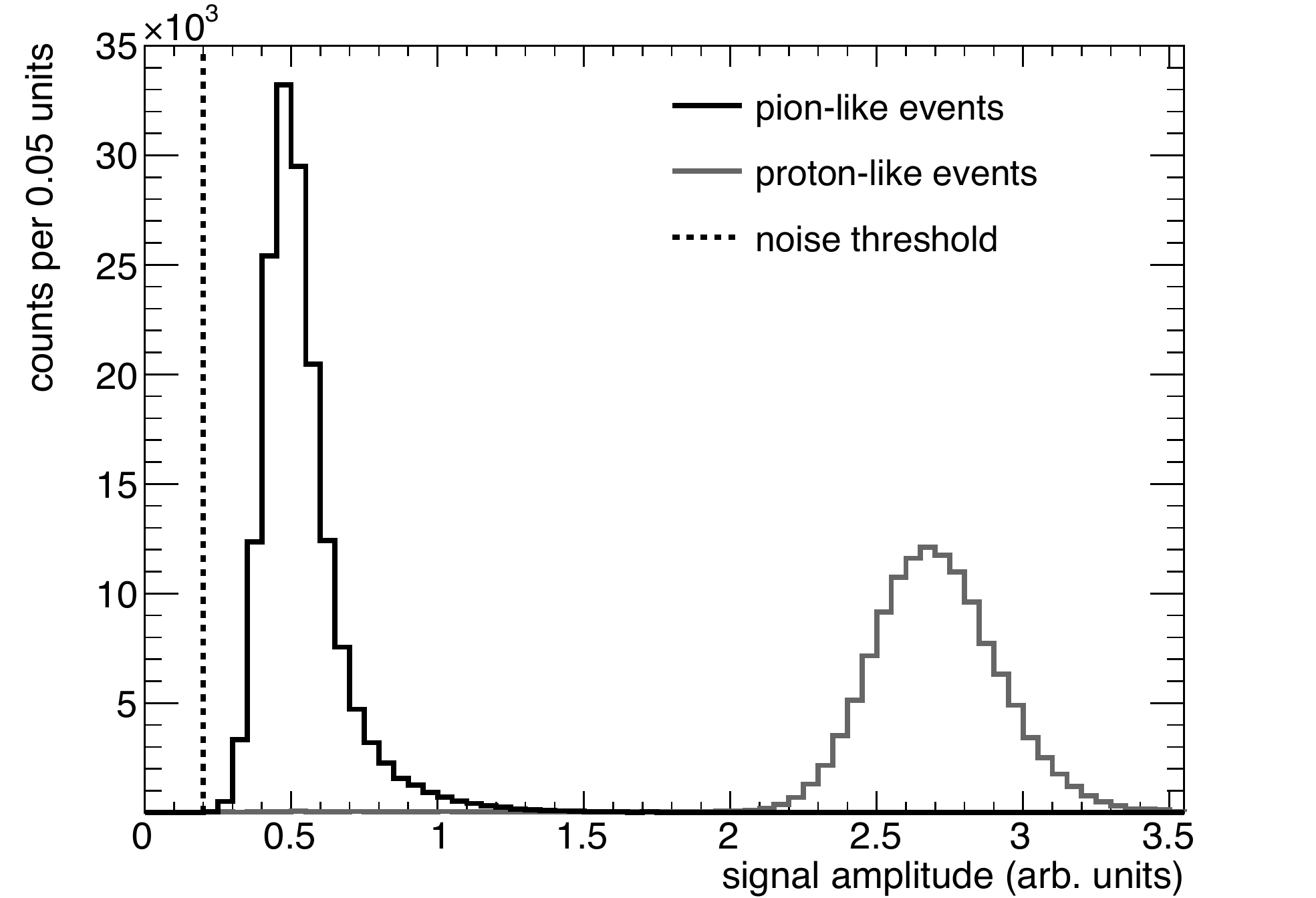}
	\caption{Pulse-height spectrum for the first fiber in the detector
    array for a beam momentum of \SI{350}{MeV/c} and an incidence
    angle of \SI{1.6}{\degree}. The fiber type is SCSF-78.
    \label{fig:single_fiber_spectrum}}
\end{figure}

In each run, the beam contains both protons and pions. An event
consists of one particle passing through a contiguous part of the
fiber array (always including the first fiber, which was used as a
trigger), producing scintillation light in each traversed fiber. The
SiPMs convert this light into charge signals, which are digitized in
the ADCs. We fit to the ADC output to determine the signal amplitude
for each fiber. Figure~\ref{fig:single_fiber_spectrum} shows the
signal-amplitude spectrum for a single fiber and a single run.

At our beam momenta, pions pass through all 16 fibers with a nearly
constant energy-loss density; accordingly we define an event as
pion-like if it passes through all fibers. The signal in each fiber
must be above the noise threshold measured for that fiber. Pion-like
events form the low-amplitude peak in
figure~~\ref{fig:single_fiber_spectrum}. The arithmetic mean of the
spectral distribution of pion-like events is the uncalibrated mean
pion light yield.

Since the minimum energy-loss density for a proton is three times
higher than that for a pion, we define a proton-like event as one with
more than one fiber with a signal amplitude exceeding three times the
mean pion light yield for that fiber. Proton-like events form the
high-amplitude peak in figure~\ref{fig:single_fiber_spectrum}. The
arithmetic mean of the spectral distribution of proton-like events is
the uncalibrated mean proton light yield. The ratio of the
uncalibrated mean proton light yield to the uncalibrated mean pion
light yield is the relative mean light yield, $\bar\Lambda_i$.

The signals from the SiPMs are smeared by noise, the resolutions of
the ADCs, the pulse-shape fits, and the event-selection algorithm. We
estimate the uncertainty on the relative mean light yield from these
effects by fitting a Landau distribution folded with a normal
distribution to the pion peak in the signal-amplitude spectrum. We
take the standard deviation of the normal distribution,
\SI{5}{\percent} (relative), as a conservative estimate of the
measurement uncertainty on the mean light yield and add it (in
quadrature) to the statistical uncertainty from the above steps. The
result is the $\sigma_{ri}$ used in the likelihood for the model fit.

\begin{figure}[t!]
	\includegraphics[width=\linewidth]{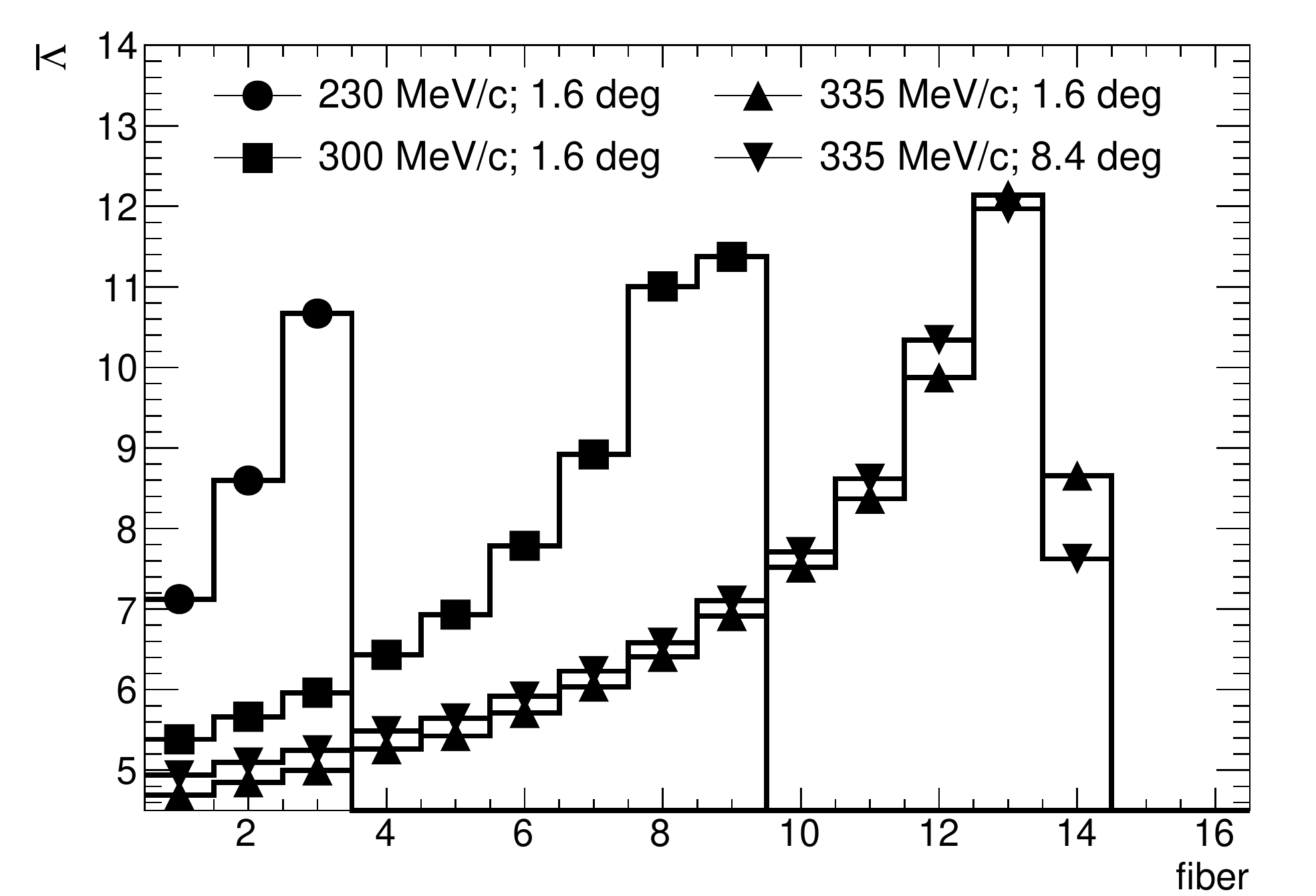}
	\caption{Relative mean light yield of proton-like events for four different runs measured with the SCSF-78 scintillator. The uncertainties are smaller than the symbols. \label{fig:BraggCurves}}
\end{figure}

Figure~\ref{fig:BraggCurves} shows the relative mean light yield of
five runs. For all four runs, we clearly see the Bragg curves for
stopping particles, with the particle range increasing with increasing
momentum. From the two runs at \SI{335}{MeV/c}, we see that changing
the incidence angle causes measurable changes in the $\bar\Lambda$
profile.

%%%%%%%%%%%%%%%%%%%%%%%%%
%%%%%%%%%%%%%%%%%%%%%%%%%
\section{Results}

To evaluate each model's posterior probability, we must choose prior
probability distributions for the model's parameters. We choose each
prior to be uniform within a reasonable range, imposing physical
constraints, and to be zero outside this range. All model parameters
are constrained by requiring
\begin{equation}
  0 \le \bar{Q}(\bar\epsilon;\vec\lambda) \le 1
  \qquad
  \forall \, \bar\epsilon \ge 0.
  \label{eqn:quenching_fn_limits}
\end{equation}
This is fulfilled for all our models when their parameters are greater
than or equal to zero. Additionally, for Voltz' model, $f$ is bounded
above by one.

\begin{table}[t!]
	\begin{center}
		\begin{tabular}{lll *{2}{ll}}
			\toprule

      model & par. & units & SCSF-78 & corr. & BC-408 & corr.\\

			\midrule\midrule

			Birks
      & $kB$
      & \si{mm/MeV}
      & \num{0.132 +- 0.004}
      &
      & \num{0.155 +- 0.005}
      &\\

 			\midrule
      
			Chou
      & $kB$
      & \si{mm/MeV}
      & $0.000 \le 0.001^{\dagger}$
      & \multirow{2}{*}{0.93}
      & \num{0.151 +- 0.04}
      & \multirow{2}{*}{0.75}
      \\
      & $\sqrt{C}$
      & \si{mm/MeV}
      & \num{0.129 +- 0.005}
      &
      & $0.000 \le 0.002^\dagger$
      & \\

 			\midrule
      
      Wright
      & $W$
      & \si{mm/MeV}
      & \num{0.333 +- 0.009}
      &
      & \num{0.406 +- 0.002}
      & \\

 			\midrule
      
      Voltz
      & $V$
      & \si{mm/MeV}
      & \num{0.091 +- 0.006}
      & \multirow{2}{*}{0.25}
      & \num{0.628 +- 0.108}
      & \multirow{2}{*}{0.89}
      \\
      & $f$
      &
      & $0.000 \le 0.057^\dagger$
      &
      & \num{0.427 +- 0.019}
      & \\

			\bottomrule
		\end{tabular}
		\caption{Parameter values at the best-fit points,
      \SI{68}{\percent}-credibility-interval uncertainties, and the
      correlation factors (where applicable). $^\dagger$These best-fit
      values are at their boundaries, zero, so we give their
      \SI{68}{\percent}-credibility upper limits.
      \label{tab:best_fit_points}}
	\end{center}
\end{table}

In table~\ref{tab:best_fit_points}, for each of the four models and
for each of the scintillating fiber types, we list the parameter
points that maximize the posterior probability, which we refer to as
the best-fit point; the \SI{68}{\percent}-credibility-interval
uncertainties; and correlation factors (where applicable). The
uncertainties and correlation factors include both statistical and
systematic effects. We are able to measure Birks' coefficient to a
relative precision of~\SI{3}{\percent}. Our value of Birks' $kB$ for
BC-408 agrees with that presented in~\cite{Almurayshid2017}.

We observe very different behavior of Chou's model for the two
scintillators: For SCSF-78, the term linear in $\bar\epsilon$ is
negligible and quenching is best described by the quadratic term
alone, with $\sqrt{C}$ compatible with Birks' $kB$. For BC-408, the
opposite is the case and quenching is best described by the linear
term alone, with Chou's $kB$ compatible with Birks'. Therefore Chou's
model requires the shape of the quenching function strongly depend on
the scintillator material.

We also observe very different behavior of Voltz' model for the two
scintillators: For SCSF-78, $f$ is small, with a best-fit value of
zero; Table~\ref{tab:best_fit_points} lists the mode and
\SI{68}{\percent}-credibility upper limit. This means that all
deposited energy is subject to quenching, as in Birks'
model. Accordingly, for this fiber type, $V$ is of a comparable scale
to Birks' $kB$. For BC-408, $f$ is closer to \SI{50}{\percent}---only
half the deposited energy is subject to quenching. Accordingly, $V$
must be larger. This trend is confirmed by the positive correlation of
the parameters in both fits with Voltz' model. Voltz' model also
requires the shape of the quenching function strongly depend on the
scintillator material.

%%%%%%%%%%%%%%%%%%%%%%%%%
\subsection{Model-independent fit}

\begin{table}[t!]
  \centering
  \begin{tabular}{l*{8}{@{\hspace{0.7em}}S}*{3}{@{\hspace{0.5em}}S}}
    \toprule
    $Q(\epsilon)$ & {998} & {996} & {962} & {922} & {856} & {629} & {419} & {405} & {309} & {113} \\%&\si{\percent} \\
    \midrule
    $\epsilon$ & {5} & {10} & {15} & {20} & {30} & {50} & {75} & {100} & {250} & {500}\\% & \si{MeV/cm} \\
	  \midrule
	  5   & 1.32 & 1.23 & 1.10 & 1.05 & 1.00 & 0.76 & 0.55 & 0.30 & 0.01 & 0.33\\% & \textperthousand \\
	  10  &      & 1.67 & 1.60 & 1.54 & 1.44 & 1.12 & 0.81 & 0.42 & 0.01 & 0.18\\% & \textperthousand \\
	  15  &      &      & 1.72 & 1.63 & 1.54 & 1.12 & 0.83 & 0.51 & 0.11 & 0.20\\% & \textperthousand \\
	  20  &      &      &      & 1.59 & 1.48 & 1.15 & 0.81 & 0.35 &-0.16 & 0.04\\% & \textperthousand \\
	  30  &      &      &      &      & 1.44 & 0.99 & 0.81 & 0.41 &-0.21 & 0.04\\% & \textperthousand \\
	  50  &      &      &      &      &      & 1.15 & 0.65 &-0.40 &-0.86 &-0.65\\% & \textperthousand \\
	  75  &      &      &      &      &      &      & 1.94 &-0.12 &-1.51 &-1.18\\% & \textperthousand \\
	  100 &      &      &      &      &      &      &      & 2.73 & 2.72 & 1.89\\% & \textperthousand \\
	  250 &      &      &      &      &      &      &      &      &\hspace{0.8em}{10.8} & 6.39\\% & \textperthousand \\
	  500 &      &      &      &      &      &      &      &      &      &\hspace{0.8em}{10.2}\\% & \textperthousand \\
    \bottomrule
  \end{tabular}
  \caption{Result of model-independent fit to SCSF-78: $Q$ values
    (top) and covariances (bottom) in \textperthousand\ at fixed
    $\epsilon$ values (in \si{MeV/cm}).
    \label{tab:SCSF78}
  }
\end{table}

\begin{table}[t!]
  \centering
  \begin{tabular}{l*{7}{@{\hspace{0.7em}}S}*{4}{@{\hspace{0.5em}}S}}
	  \toprule
    $Q(\epsilon)$ & {989} & {847} & {768} & {709} & {652} & {517} & {444} & {412} & {263} & {115} \\% & \si{\percent} \\
    \midrule
	  $\epsilon$ & {5} & {10} & {15} & {20} & {30} & {50} & {75} & {100} & {250} & {500} \\% & \si{MeV/cm}\\
	  \midrule
	  5   & \num{2.32} & 1.43 & 1.09 & 1.01 & 1.01 & 0.45 & 0.55 & 0.46 &  0.50 & 0.28 \\% & \textperthousand \\
	  10  &            & 3.88 & 3.95 & 3.67 & 3.59 & 2.32 & 1.23 & 0.73 &  0.94 & 0.69 \\% & \textperthousand \\
	  15  &            &      & 4.22 & 3.90 & 3.89 & 2.22 & 1.12 & 0.61 &  0.80 & 0.63 \\% & \textperthousand \\
	  20  &            &      &      & 3.65 & 3.65 & 2.06 & 0.76 & 0.23 &  0.50 & 0.42 \\% & \textperthousand \\
	  30  &            &      &      &      & 4.06 & 0.85 &-0.41 &-8.20 & -0.35 &-0.10 \\% & \textperthousand \\
	  50  &            &      &      &      &      & 6.10 & 3.33 & 1.22 &  1.15 & 0.91 \\% & \textperthousand \\
	  75  &            &      &      &      &      &      & 7.94 & 6.65 &  4.31 & 2.81 \\% & \textperthousand \\
	  100 &            &      &      &      &      &      &      &\hspace{0.8em}{10.1}&  7.23 & 3.90 \\% & \textperthousand \\
	  250 &            &      &      &      &      &      &      &      &  8.17 & 4.66 \\% & \textperthousand \\
	  500 &            &      &      &      &      &      &      &      &       & 5.72 \\% & \textperthousand \\
	  \bottomrule
  \end{tabular}
  \caption{Result of model-independent fit to BC-408: $Q$ values
    (top) and covariances (bottom) in \textperthousand\ at fixed
    $\epsilon$ values (in \si{MeV/cm}).
    \label{tab:BC408}
  }
\end{table}

The models we tested impose strong assumptions on the form of
$Q(\epsilon)$: all but Chou's model have positive second derivatives
for all $\epsilon$; all but Voltz' model approach zero at large
$\epsilon$. Since a material's quenching function has never been
directly measured before, these assumptions have gone untested. The
data collected with our segmented detector allows us to directly fit
for the shape of the quenching function free from model
assumptions. For this, we parametrized $Q(\epsilon)$ as a
linear spline with eleven knots. We tried several different model-independent descriptions: using a cubic spline instead of a linear one; freeing the knot positions in the fit; and using fewer or more knots. The results were all consistent with each other. We show the results with fixed knot positions and a linear interpolation because it is the simplest to present the full results for, including parameter correlations. The knots positions were fixed at 0, 5, 10, 15, 20, 30, 50, 75, 100, 250, and 500~\si{MeV/cm}. We chose these values to cover the full range of $\epsilon$ of our experiment and have a higher knot density in regions our experiment is most sensitive to. The value of the quenching function at $\epsilon=\SI{0}{MeV/cm}$ is fixed to unity. For
our model-independent fits we used uniform prior probabilities on the
value of $Q$ at each knot. The resulting best-fit values and
covariances are listed in tables~\ref{tab:SCSF78} and~\ref{tab:BC408}.

\begin{figure*}
	\begin{subfigure}[b]{1.\linewidth}
		\includegraphics[width=\linewidth]{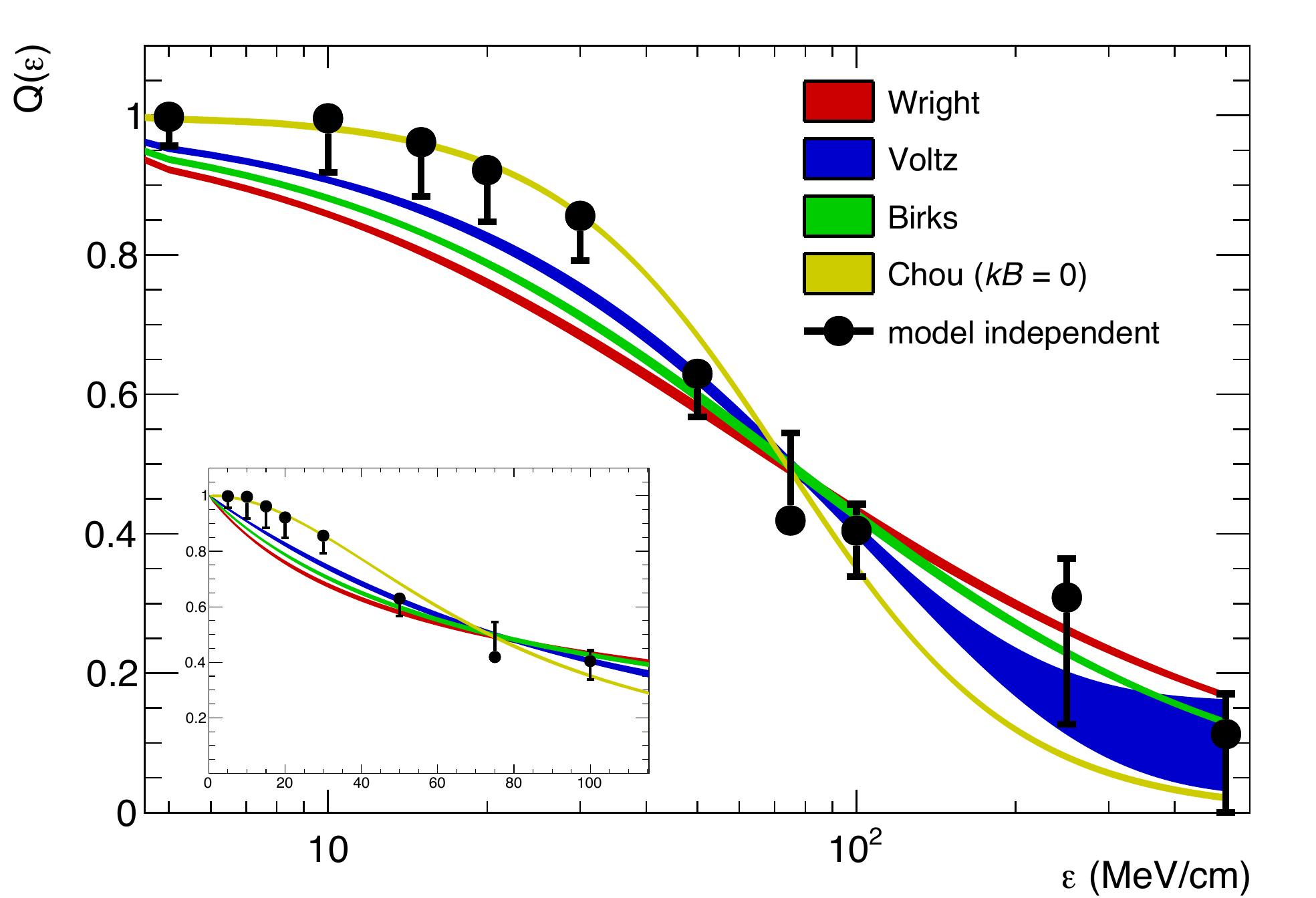}
		\caption{SCSF-78.}
		\label{fig:dLdESCSF78}
	\end{subfigure}
	\begin{subfigure}[b]{1.\linewidth}
		\includegraphics[width=\linewidth]{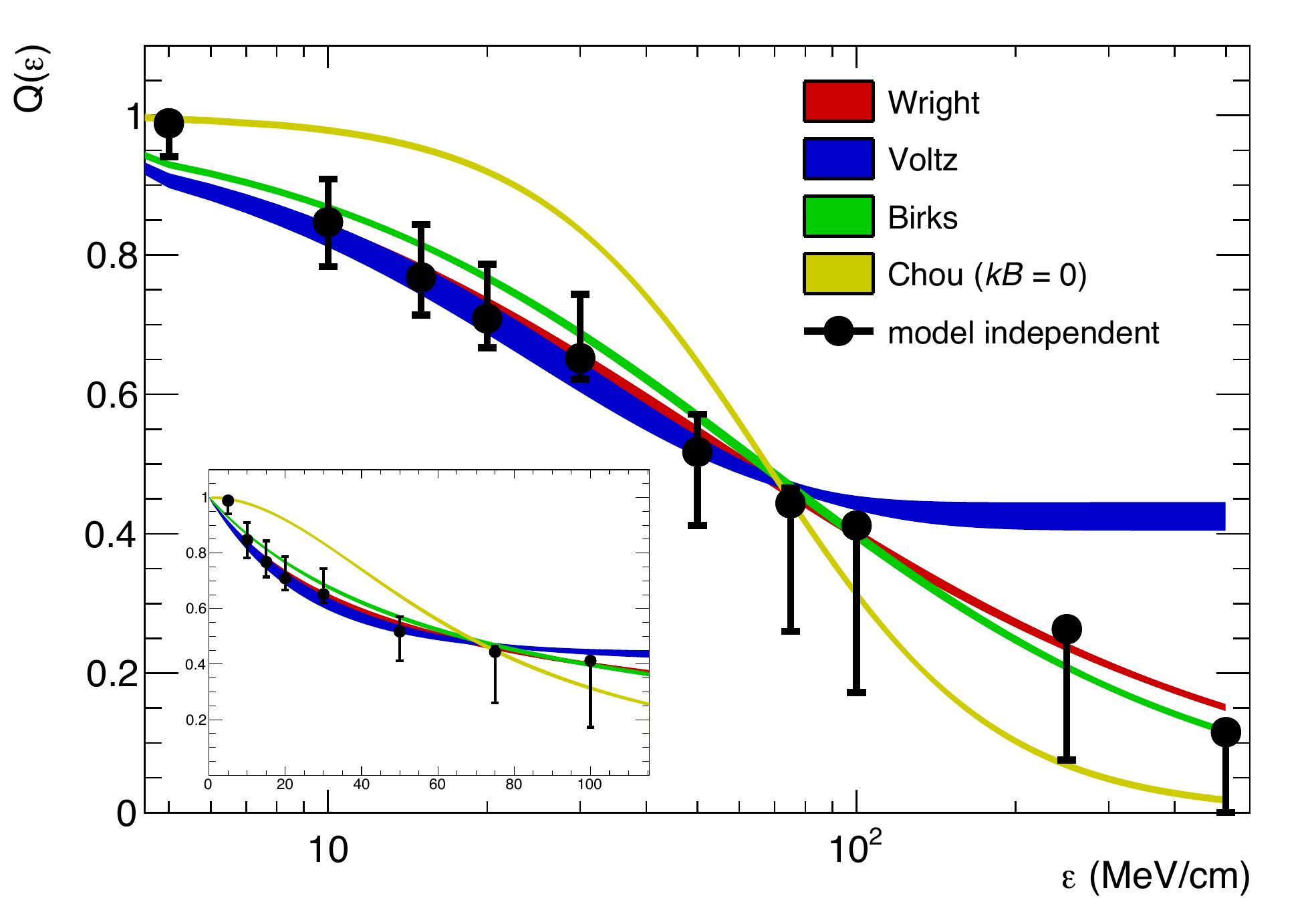}
		\caption{BC-408.}
		\label{fig:dLdEBC408}
	\end{subfigure}
	\caption{\SI{68}{\percent}-credibility-interval bands for the
    model-dependent quenching functions listed in
    table~\ref{tab:best_fit_points} and the best-fit values and the
    smallest \SI{68}{\percent}-credibility intervals for the
    model-independent fits for both scintillator materials. The inset
    plots show the same results on a linear scale for the
    low-$\epsilon$ region; their axes have the same variables and
    units as the larger plots. \label{fig:quenching_functions}}
\end{figure*}

Figure~\ref{fig:quenching_functions} shows the
\SI{68}{\percent}-credibility-interval bands for the model-dependent
quenching functions and the best-fit values for the model-independent
quenching functions for both scintillators. The bars on the
model-independent results show the boundaries of the smallest
\SI{68}{\percent}-credibility intervals for the value at each
knot. For many of the knots, the best-fit value is near the boundary
of the interval---especially those near unity. The results of the
model-independent fits yield quenching functions free from any
theoretically-imposed constraints. Using these results, we
qualitatively evaluate each model's ability to reproduce the data.

The model-independent quenching function for SCSF-78 has a negative
second derivative at small $\epsilon$ and an inflection point at
approximately \SI{50}{MeV/cm}. It is inconclusive whether the
quenching function approaches zero at large $\epsilon$---the value of
the quenching function at \SI{500}{MeV/cm} is \num{1.5} standard
deviations above zero (in the posterior probability). Only Chou's
model, when $C$ is nonzero, can accommodate a negative second
derivative. Figure~\ref{fig:dLdESCSF78} includes the result of a
fit using Chou's model with $kB$ fixed to zero, which is identical to
the fit result reported in table~\ref{tab:best_fit_points}. We see
that Chou's model is able to describe the small-$\epsilon$ behavior
better than all other models.

The model-independent quenching function for BC-408 has a positive
second derivative everywhere. It again is inconclusive whether it tends to
zero or to a finite quenching value at large $\epsilon$---the value of
the quenching function at \SI{500}{MeV/cm} is \num{1.1} standard
deviations above zero (in the posterior probability). These features
are compatible with all the model-dependent fits. Figure~\ref{fig:dLdEBC408} includes the result of a fit using Chou's
model with $kB$ fixed to zero---we do not show the result for a free
$kB$ since it is identical to the fit with Birks' model. We conclude
that this model cannot describe the data well because it must have a
negative second derivative at small $\epsilon$, which is contradicted
by the model-independent result.

The model-independent quenching functions indicate that it is likely
that quenching does not asymptotically drop to zero and light is
produced even at large energy-deposition density.

%%%%%%%%%%%%%%%%%%%%%%%%%
\subsection{Model comparisons}

\begin{table}[t!]
	\begin{center}
		\begin{tabular}{l*{4}{S}}
			\toprule

      & \multicolumn{2}{l}{SCSF-78} & \multicolumn{2}{l}{BC-408} \\
      \cmidrule(lr){2-3} \cmidrule(lr){4-5}
      
			model
      & {$\log_{10} K$}
      & {$\Delta\log_{10} L$}
      & {$\log_{10} K$}
      & {$\Delta\log_{10} L$}\\

			\midrule
      
			Birks
      & {---} & {---} & {---} & {---}
      \\ 

 			Chou
      & 5.1 +- 0.2
      & 14.6
      & -3.6 +- 0.1
      & 0.2
      \\

      Chou ($kB=0$)
      & 7.9 +- 0.2
      & 15.0
      & -14.7 \pm 0.1
      & -21.3
      \\

      Wright
	    & -19.0 +- 0.2
      & -3.0
      &	1.0 +- 0.1
      & 0.1
      \\

      Voltz
      & 5.9 +- 0.2
      & 5.6
      & 1.0 +- 0.1
	    & 1.1
	    \\

      % Logistic
      % & 12.0 +- 0.5
      % & {???}
      % & 6.4 +- 0.2
      % & {???}
      % \\
      
			\bottomrule
		\end{tabular}
		\caption{The log of the Bayes factor and the difference of the log
      of the maximum likelihood, both with respect to the fit for
      Birks' model, for both scintillators.
		\label{tab:model_comparison}}
	\end{center}
\end{table}

Table~\ref{tab:model_comparison} compares our fits for each model for
both scintillators: we list both Bayes factors and the difference in
maximum likelihood.\footnote{Though there is no simple, true
  statistical interpretation of the difference in maximum likelihood
  as a basis for model comparison, we give this information since it
  is commonly used in the field.} We benchmark all models against
Birks' model, which is the most commonly used quenching model. If a
model fits to the data better than Birks' model, the value of
$\log_{10} K$ is positive; if a model fits to the data worse than
Birks' model, it is negative. Common interpretations of Bayes factors
state that $\abs{\log_{10} K} > 2$ means there is decisive evidence
for a conclusion; and $\abs{\log_{10} K} \approx 1$ means there is
only substantial evidence~\cite{Kass:1995loi, Jeffreys:2003}.

Our fits to the SCSF-78 data decisively prefer Chou's and Voltz'
models to Birks', with no strong evidence for a preference of either
one over the other. However, Chou's model with $kB$ fixed to zero is
decisively preferred to all other models---its preference over Chou's
full model is a clear example of Occam's razor. Wright's model is
strongly disfavored by our data.

These conclusions are borne out in visual comparison to the
model-independent functions (figure~\ref{fig:dLdESCSF78}): Chou's
model with $kB = 0$ is the only model that reproduces the
model-independent function for SCSF-78 at small $\epsilon$. Our fits
are most sensitive to behavior at small $\epsilon$, where a
preponderance of our data is. So Chou's model with $kB=0$ is still
preferred to the other models, though it deviates the most from the
model-independent behavior at medium and large $\epsilon$. To better
study the behavior at large $\epsilon$, we need data using heavier and
higher-charged particles, namely ions.

In fits to the BC-408 data, Birks' model is decisively preferred to
Chou's model. This is expected: the fit with Chou's model prefers
$\sqrt{C} = 0$, recreating Birks' model but with an extra degree of
freedom. This unnecessary degree of freedom is a penalty when
calculating the Bayes factor---again an example of Occam's
razor. Wright's and Voltz' models are substantially preferred. The
Bayes factor for comparing Wright's model (with its evidence in the
numerator) to Voltz' (in the denominator) is \num{0.1\pm0.1}, barely
favoring Wright's model, but inconclusively. That none of Birks',
Voltz', or Wright's models is decisively preferred, is also borne out
in visual comparison to the model-independent function
(figure~\ref{fig:dLdEBC408}): all three models reproduce the
model-independent results within their \SI{68}{\percent} credibility
intervals.

Our studies above show that quenching in SCSF-78 and in BC-408 have
different dependencies on energy-deposition density. The two
scintillator types differ in base material, dopant material, and
dopant density---all of which can contribute to differences in
quenching. No model we tested is decisively favored in fits with both
scintillators. Chou's model with $kB=0$ is most favored in fits to
SCSF-78 data, but most disfavored in fits to BC-408 data. The only
model to perform better than Birks' in both fits is Voltz'.

A new model is needed to parametrize quenching in both materials. The
most generic model that could fit all the features seen in the
model-independent fits must allow for an asymptotic value at large
$\epsilon$; the possibility of a negative second derivative at small
$\epsilon$ with an inflection point where the second derivative may
change sign; and different curvatures below and above this inflection
point. Such a model would require at least four parameters, with all
or some of them being specific to the material composition used. To
fully test such a model requires new measurements at small, medium,
and large $\epsilon$ for multiple scintillating materials.

%%%%%%%%%%%%%%%%%%%%%%%%%
%%%%%%%%%%%%%%%%%%%%%%%%%
\section{Conclusion}

We have reported measurements to precisely determine the light yield
dependence on the energy-deposition density by charged particles for
two different scintillating materials and presented a novel method of
fitting quenching functions to this data. We have determined the
parameters of four widely used quenching models---Birks', Chou's,
Wright's, and Voltz'---with percent-level precision. This is the first
report of these parameters for the SCSF-78 scintillator; and the first
report of the parameters for Chou's, Wright's, and Voltz' models for
the BC-408 scintillator.

We have also determined the dependence of ionization quenching on
energy-deposition density for both scintillating materials using a
model-independent technique. To our knowledge, this is the first
model-independent determination of quenching functions. Our results
indicate that quenching is highly dependent upon the scintillating
material, with no common model strongly preferred for both materials;
none of the most-commonly-used models describe the features of the
true quenching function over the full range of energy-loss density
well. Owing to their assumptions on the shape of the quenching
function, these models will always overestimate quenching at either
low or high energy-deposition density. The constraints of the models
and that most quenching measurements are made over a small range of
energy-deposition density explains why various measurements with the
same scintillator type often do not agree with each
other~\cite{Craun1970,Hirschberg1992}. Our model-independent approach
allows us to determine quenching over a large range of
energy-deposition densities without subregions biasing each other. New
and more refined models can now be developed using our
model-independent quenching functions.

%%%%%%%%%%%%%%%%%%%%%%%%%
%%%%%%%%%%%%%%%%%%%%%%%%%
\section{Acknowledgments}

We would like to thank O.~Schulz and R.~Schick for providing their HME
code; P.~von~Doetinchem for providing the BC-408 scintillator;
I.~Konorov, S.~Huber, and D.~Levit for their help with data
acquisition; K.~Deiters, T.~Rauber, and M.~Schwarz for their support
prior and during the experiment at Paul Scherrer Institute; and our
late colleague D.~Renker for his persistent support and many fruitful
discussions.

This research was supported by the DFG Cluster of Excellence
\emph{Origin and Structure of the Universe} (EXC 153).

%%%%%%%%%%%%%%%%%%%%%%%%%
%%%%%%%%%%%%%%%%%%%%%%%%%

\bibliography{quenching_paper_references}

\end{document}